# Dynamical polarization of nuclear spins by acceptor-bound holes in a zinc blende semiconductor

K.V. Kavokin and A.V. Koudinov

A.F. Ioffe Physico-Technical Institute of RAS, 194021 St.-Petersburg, Russia

and

Spin Optics Laboratory, St.-Petersburg State University, 198504 St.-Petersburg, Russia

The ground state of an acceptor-bound hole in a zinc-blende semiconductor is formed by four eigenstates of the total angular momentum, which is a vector sum of spin and orbital moment of the hole. As a result, the hyperfine interaction of the hole with lattice nuclei becomes anisotropic and coordinate-dependent. We develop a theory of dynamic polarization of nuclear spins by the acceptor-bound hole, giving full account for its complex spin structure. The rate of hole-nuclear flip-flop transitions is shown to depend on the angle between the total angular momentum of the hole and the position vector of the nucleus with respect to the acceptor center. The resulted spatially inhomogeneous spin polarization of nuclei gives rise to non-equidistant spin splitting of the hole, which can be detected by methods of optical or microwave spectroscopy.

## 1. Introduction

The dynamical polarization of spins of the nuclei which constitute the crystal lattice of a semiconductor has been known since the pioneering paper by Lampel.[1] As a rule, it originates from the Fermi contact interaction binding the nuclear spins with spins of the conduction-band electrons whose wave functions are formed by the *s*-type atomic orbitals. Holes in the valence band, formed usually by the p-type orbitals, also interact with the nuclear spins. However this interaction is about one order of magnitude weaker, since the density of the *p*-electron wave function turns to zero at a position of the nucleus.[2–4] This circumstance, together with the strong spin-orbit coupling in the valence band leading to fast spin relaxation of the free holes, substantially weaken the effects of the hole-nuclear interaction. Until recently, these effects were usually neglected against the background of the pronounced electron effects. In the recent papers[5–7] the hyperfine splitting of the localized states of holes in semiconductor quantum dots was experimentally measured and theoretically calculated. The role of holes in the dynamical polarization of nuclei in quantum dots was also theoretically considered.[8] Note that as a rule, the light-hole states in quantum dots are split off because of both the confinement and the strain, so that the ground state of the hole is the Kramers doublet formed mainly by the heavy-hole subband states.

The holes localized by a spherically-symmetric potential (e.g., on an acceptor center or in a spherical nanocrystal) possess a much more complex and interesting spin structure.[9–13] The ground state is fourfold degenerate, and its basis functions are the eigenfunctions of the total angular momentum formed by the vector summation of the spin and the orbital moment of the localized hole. As a result, the matrix elements and the average value of the spin moment operator become coordinate-dependent. Manifestations of the complex structure of the acceptor center in the spin-spin interactions were theoretically studied for the case of a diluted magnetic semiconductor, where they lead to a non-trivial geometry of the spin ordering of localized spin moments within the magnetic polaron.[14,15]

In the present paper, the problem of dynamical polarization of nuclei by the acceptor-bound hole is treated theoretically. The rate of pumping of nuclear spins and the steady-state polarization of the nuclei is calculated as a function of coordinates in the vicinity of the acceptor center in a GaAs-type semiconductor. The splitting of the four ground-state sublevels of the acceptor, induced by the interaction with the dynamically-polarized nuclei, is shown to be non-equidistant. This makes the notion of the effective Overhauser field not quite applicable. Possible experiments aimed at the observation of the effect are discussed.

## 2. General considerations

### 2.1 Hyperfine interaction of the hole

In the general form, the Hamiltonian of the spin-spin interaction between the hole with the angular momentum $J=3/2$ and a localized spin (particularly, the nuclear spin $\vec{I}$) can be written as[15]

$$\tilde{H}_S = a(\rho)(\vec{J}\cdot\vec{I}) + b(\rho)(\vec{J}\cdot\vec{\rho})^2(\vec{J}\cdot\vec{I}) + c(\rho)(\vec{J}\cdot\vec{\rho})(\vec{\rho}\cdot\vec{I}) + d(\rho)(\vec{J}\cdot\vec{\rho})^3(\vec{\rho}\cdot\vec{I}), \tag{1}$$

where $\vec{\rho}$ is the position vector of the nucleus with respect to the localization center of the valence-band orbital.

As shown in Ref. 4, the hyperfine coupling of the hole it determined, to within a 1% precision, by the interaction with the nucleus at which the atomic orbitals of the valence-band electron are centered. For this interaction, the position vector $\vec{\rho}$ equals zero, and only the first term in Eq.(1) remains:

$$\tilde{H}_{SC} = a(\vec{J}\cdot\vec{I}). \tag{2}$$

Hence in this approximation, the hyperfine coupling of the hole takes a scalar form, no matter it has a magneto-dipole origin. The interaction constant $a$ can be found by a direct calculation of any matrix element of the Hamiltonian of the magneto-dipole interaction with the Bloch functions of the valence band. As shown in Appendix A, it is equal to

$$a = \frac{64}{45}\pi\frac{\mu_B \mu_I}{I}\wp, \tag{3}$$

where $\mu_B$ and $\mu_I$ stand for the Bohr magneton and the nuclear magneton, respectively, $\wp = \int_0^\infty u^2(\rho)\rho d\rho$, and $u(\rho)$ denotes the radial part of the Bloch function of the valence band electron.

## 2.2 Spin structure of the acceptor and dynamical polarization of nuclei

Due to the strong spin-orbit interaction in the valence band, neither the spin $\vec{J}$ nor the orbital moment $\vec{L}$ of the acceptor-bound hole are conserving values. In the spherical approximation (i.e., neglecting terms of the fourth power in the wave vector, which cause the cubic-symmetry corrugation of the valence band), the acceptor ground state, being degenerate at zero magnetic field, corresponds to a definite value of the total angular momentum $\vec{F} = \vec{J} + \vec{L}$ (usually $F = 3/2$), while its basis functions correspond to definite projections of $\vec{F}$, $M = F_Z$.[9,10] If an external force (e.g., the optical pumping) drives the distribution of acceptors over the directions of the vector $\vec{F}$ off the equilibrium, the hyperfine interaction leads to the transfer of angular momentum and energy into the nuclear spin system, causing the dynamical polarization of the nuclear spins.[16,17] Following Ch.5 of Ref. 16, we consider the vector $\vec{j}(\vec{r})$ which describes the flux of the angular momentum into the nuclear spin system at a position $\vec{r}$. We shall restrict our consideration to the case of weak spin polarizations, for which the hole spin density matrix $\hat{P}$ is fully determined by the vector $\langle\delta\vec{F}\rangle = \langle\vec{F} - \vec{F}_T\rangle$, where $\vec{F}_T$ is the thermodynamically equilibrium value of the total moment of the acceptor, so that

$$P_{MM'} \approx \frac{1}{2F+1}\left[\delta_{MM'} + \frac{3}{F(F+1)}\left(\langle\delta F\rangle_x \hat{F}^X_{MM'} + \langle\delta F\rangle_y \hat{F}^Y_{MM'} + \langle\delta F\rangle_z \hat{F}^Z_{MM'}\right)\right] =$$
$$\frac{1}{4}\left[\delta_{MM'} + \frac{4}{5}\left(\langle\delta F\rangle_x \hat{F}^X_{MM'} + \langle\delta F\rangle_y \hat{F}^Y_{MM'} + \langle\delta F\rangle_z \hat{F}^Z_{MM'}\right)\right]$$

(4)

In this case, the $\vec{j}(\vec{r})$ should be linear in $\langle\delta\vec{F}\rangle$. Due to the spherical symmetry of the problem and the time-reversal symmetry, this relation has the following general form:

$$\vec{j} = f_1(r,B)\langle\delta\vec{F}\rangle + f_2(r,B)\left(\vec{r}\cdot\langle\delta\vec{F}\rangle\right)\vec{r} + f_3(r,B)\left(\vec{B}\cdot\delta\vec{F}\right)\vec{B} + f_4(r,B)\left(\vec{r}\cdot\delta\vec{F}\right)\left(\vec{r}\cdot\vec{B}\right)\vec{B} + f_5(r,B)\left(\vec{B}\cdot\delta\vec{F}\right)\left(\vec{r}\cdot\vec{B}\right)\vec{r}$$

(5)

To obtain a considerable spin polarization of nuclei, the decay of the non-equilibrium nuclear spin on a timescale of the nuclear spin-spin interactions $T_2$ must be suppressed. This is achieved by application of an external magnetic field, converting the flux of angular momentum into the flux of energy.[16,17] In what follows, we shall consider the simplest arrangement where the non-equilibrium angular momentum of the hole is directed along the external field, while this latter field is stronger than the local magnetic fields the nuclei produce on each other. Then the spin component directed along the external magnetic field is introduced into the nuclear system and persists for a relatively long spin-lattice relaxation time $T_1$ (which is determined, at low temperatures, by the same interaction with the hole). The remaining spin components quickly vanish, so that their average values are negligible. Thus the only important component of the spin flux will be that parallel to the magnetic field (whose direction is a natural choice of the $Z$ axis). Using Eq.(5), one can express it in the following form:

$$j_Z = \mathcal{A}(r,B)\langle\delta F\rangle + \mathcal{B}(r,B)\langle\delta F\rangle r^2 \cos^2\theta, \qquad (6)$$

where $\langle\delta F\rangle$ is the length of the vector $\langle\vec{\delta F}\rangle$ directed along $Z$, while $\theta$ is the angle between the $Z$ axis and the position vector $\vec{r}$. The new functions $\mathcal{A}$ and $\mathcal{B}$ are defined as $\mathcal{A}(r,B) = f_1(r,B) + f_3(r,B)B^2$ and $\mathcal{B}(r,B) = f_2(r,B) + [f_4(r,B) + f_5(r,B)]B^2$.

Similar considerations allow one to obtain the expression for the spin flux related to the relaxation of the nuclear magnetization on the acceptor-bound hole:

$$\tilde{j}_Z = \tilde{\mathcal{A}}(r,B)\langle I_Z(r,\theta)\rangle + \tilde{\mathcal{B}}(r,B)\langle I_Z(r,\theta)\rangle r^2 \cos^2\theta, \qquad (7)$$

where $\langle I_Z(r,\theta)\rangle$ stands for the mean $Z$-projection of the nuclear spins at a point specified by spherical coordinates $(r,\theta)$.

Therefore, seeking for a coordinate dependence of $j_z(\vec{r})$, it is sufficient to calculate, at the given magnetic field, two functions of the distance from the acceptor center, $\mathcal{A}(r,B)$ and $\mathcal{B}(r,B)$. One can perform it, e.g., by calculating $j_z(\vec{r})$ at $\vec{r} \parallel \vec{B}$ and $\vec{r} \perp \vec{B}$. The same is true for $\tilde{j}_z(\vec{r})$, with functions $\tilde{\mathcal{A}}(r,B)$ and $\tilde{\mathcal{B}}(r,B)$.

## 3. The steady state of the optically-pumped system of the acceptor and nuclei

### 3.1 Calculation of the generation and relaxation spin fluxes

We shall assume throughout that the correlation time of the hole angular momentum, $\tau_c$, determined by photoexcitation and/or interaction with phonons, is short ($\sum_i \langle V^2_{iMM'mm'}\rangle \tau_c^2 / \hbar^2 \ll 1$, where $V_{iMM'}$ stands for the matrix element of the transition between hole spin sublevels $M$ and $M'$ due to hyperfine interaction with the i-th nucleus, angular brackets denote averaging over the initial spin states of nuclei, $m$). This condition means that the hole spin state is not considerably changed by the action of the fluctuating effective fields of nuclei during the correlation time, and therefore the rate of transitions with the spin-flip of a certain nucleus is not affected by the hole's interaction with the other nuclei. The above assumption, which is often valid for donor-bound electrons,[18] is even more justified for holes due to their weaker hyperfine coupling. Within this approximation of short $\tau_c$, the rate of transitions between hole sublevels $M$ and $M'$, accompanied by the transition between states $m$ and $m'$ of $i$-th nucleus, as a function of the magnetic field can be calculated as[16]

$$w_{iMM'mm'} = \frac{\hbar^{-2} V_{iMM'mm'}^2 \tau_c}{1 + \gamma_h^2 (M - M')^2 B^2 \tau_c^2} \quad (8)$$

where $\gamma_h$ is the gyromagnetic ratio of the hole. Depending on the strength of the external magnetic field, one can consider two distinct regimes of dynamic nuclear polarization. In the weak-field regime ($B\gamma_h \tau_c \ll 1$), the transition rate is field-independent and proportional to the correlation time:

$$w_{iMM'mm'} = \hbar^{-2} V_{iMM'mm'}^2 \tau_c . \quad (9)$$

In the strong-field regime ($B\gamma_h \tau_c \gg 1$), the transition rate is inversely proportional to the correlation time (i.e. directly proportional to the homogeneous width of the hole sublevels $2\pi\hbar/\tau_c$):

$$w_{iMM'mm'} = \hbar^{-2} V_{iMM'mm'}^2 \gamma_h^{-2} (M - M')^{-2} B^{-2} \tau_c^{-1} . \quad (10)$$

Kinetics of the mean Z-projection of $i$-th nuclear spin is given by the equation

$$\frac{d\langle I_z^i \rangle}{dt} = \sum_{M,M',m,m'} w_{iMM'mm'} p_M^h p_m^N (m'-m), \quad (11)$$

where $p_M^h$ and $p_m^N$ are occupation probabilities of hole and nuclear spin states, correspondingly, while the rates $w_{iMM'mm'}$ are determined by either Eq. (9) or Eq. (10), depending on the external field strength. Using the hole spin density matrix given by Eq. (4), and assuming analogous statistical distribution of the nuclear spin, we get

$$\frac{d\langle I_z^i \rangle}{dt} = \frac{1}{2F+1} \frac{1}{2I+1} \sum_{M,M',m,m'} w_{iMM'mm'} \left(1 + \frac{3\langle \delta F \rangle}{F(F+1)} M\right)\left(1 + \frac{3\langle I_z \rangle}{I(I+1)} m\right)(m'-m) \approx$$
$$\frac{1}{2F+1} \frac{1}{2I+1} \sum_{M,M',m,m'} w_{iMM'mm'} \left(\frac{3\langle \delta F \rangle}{F(F+1)} M + \frac{3\langle I_z \rangle}{I(I+1)} m\right)(m'-m) \quad (12)$$

Here we made use of smallness of $\langle \delta F \rangle$ and $\langle I_z \rangle$ as well as of the relation $w_{iMM'mm'} = w_{iM'Mm'm}$.

As shown in previous section, due to the symmetry of the problem it is sufficient to calculate the generation and relaxation of the mean nuclear spin along two directions: along $\langle \delta \vec{F} \rangle$ and perpendicular to it. The complete wave functions of the acceptor ground state with the angular momentum projections $M = \pm 3/2, \pm 1/2$ are given in Appendix B. The radial dependence of the envelope functions is determined by combinations of functions $R_0 = R_0(R)$ and $R_2 = R_2(R)$, while their angular dependence – by spherical harmonics. In the longitudinal direction, $\vec{R} \| \langle \delta \vec{F} \rangle$ ($\theta = 0$), the latter are reduced to: $Y_{0,0} = 1/\sqrt{4\pi}$, $Y_{2,0} = -\sqrt{5/4\pi}$, $Y_{2,\pm 1} = Y_{2,\pm 2} = 0$. So, one obtains from Eqs. (B1)

$$\psi_{\pm 3/2} = \frac{1}{\sqrt{4\pi}} u_{\pm 3/2} R_-, \quad \psi_{\pm 1/2} = \frac{1}{\sqrt{4\pi}} u_{\pm 1/2} R_+, \tag{13}$$

where $R_+ = R_0 + R_2$, $R_- = R_0 - R_2$, and the Bloch functions $u_{\pm 3/2}$, $u_{\pm 1/2}$ are defined in Appendix A.

The interaction of holes with nuclei at the primitive-cell level is considered in detail in Appendix A. Using Eq. (A9) (which is the matrix representation of the general spin Hamiltonian Eq. (2) in the basis of the valence-band Bloch functions), one immediately obtains the matrix elements of the hyperfine interaction calculated with the acceptor functions Eqs. (13):

$$\langle \Psi_M | \hat{H}_{dd} | \Psi_{M'} \rangle = \frac{8}{15} \frac{\mu_B \mu_I}{I} \wp \begin{pmatrix} R_-^2 \hat{I}_z & \frac{R_+ R_-}{\sqrt{3}} \hat{I}_- & 0 & 0 \\ \frac{R_+ R_-}{\sqrt{3}} \hat{I}_+ & \frac{R_+^2}{3} \hat{I}_z & \frac{2}{3} R_+^2 \hat{I}_- & 0 \\ 0 & \frac{2}{3} R_+^2 \hat{I}_+ & -\frac{R_+^2}{3} \hat{I}_z & \frac{R_+ R_-}{\sqrt{3}} \hat{I}_- \\ 0 & 0 & \frac{R_+ R_-}{\sqrt{3}} \hat{I}_+ & -R_-^2 \hat{I}_z \end{pmatrix}. \tag{14}$$

In the transverse direction, $\vec{R} \perp \langle \delta \vec{F} \rangle$ ($\theta = \pi/2$, and choosing $\phi = 0$ without loss of generality), the spherical harmonics reduce to $Y_{0,0} = 1/\sqrt{4\pi}$, $Y_{2,0} = -\sqrt{5/16\pi}$, $Y_{2,\pm 1} = 0$, $Y_{2,\pm 2} = -\sqrt{5/32\pi}$. Now each of the acceptor wave functions is a linear combination of two Bloch amplitudes:

$$\psi_{\pm 3/2} = \frac{1}{\sqrt{4\pi}} \left( \tilde{R}_+ u_{\pm 3/2} - \frac{\sqrt{3}}{2} R_2 u_{\mp 1/2} \right), \quad \psi_{\pm 1/2} = \frac{1}{\sqrt{4\pi}} \left( \tilde{R}_- u_{\pm 1/2} - \frac{\sqrt{3}}{2} R_2 u_{\mp 3/2} \right), \tag{15}$$

where $\tilde{R}_+ = R_0 + R_2/2$, $\tilde{R}_- = R_0 - R_2/2$. In the end, instead of Eq. (14), one obtains

$$\langle \Psi_M | \hat{H}_{dd} | \Psi_{M'} \rangle = \frac{8}{15} \frac{\mu_B \mu_I}{I} \wp \begin{pmatrix} \left( \tilde{R}_+^2 - \frac{R_2^2}{4} \right) \hat{I}_z & \frac{R_0^2}{\sqrt{3}} \hat{I}_- - \frac{\tilde{R}_- R_2}{\sqrt{3}} \hat{I}_+ & -\frac{R_+ R_2}{\sqrt{3}} \hat{I}_z & \frac{R_2^2}{2} \hat{I}_- - \tilde{R}_+ R_2 \hat{I}_- \\ \frac{R_0^2}{\sqrt{3}} \hat{I}_+ - \frac{\tilde{R}_- R_2}{\sqrt{3}} \hat{I}_- & \left( \frac{\tilde{R}_-^2}{3} - \frac{3 R_2^2}{4} \right) \hat{I}_z & \frac{2}{3} \tilde{R}_-^2 \hat{I}_- - \tilde{R}_- R_2 \hat{I}_+ & \frac{R_+ R_2}{\sqrt{3}} \hat{I}_z \\ -\frac{R_+ R_2}{\sqrt{3}} \hat{I}_z & \frac{2}{3} \tilde{R}_-^2 \hat{I}_+ - \tilde{R}_- R_2 \hat{I}_- & -\left( \frac{\tilde{R}_-^2}{3} - \frac{3 R_2^2}{4} \right) \hat{I}_z & \frac{R_0^2}{\sqrt{3}} \hat{I}_- - \frac{\tilde{R}_- R_2}{\sqrt{3}} \hat{I}_+ \\ \frac{R_2^2}{2} \hat{I}_- - \tilde{R}_+ R_2 \hat{I}_+ & \frac{R_+ R_2}{\sqrt{3}} \hat{I}_z & \frac{R_0^2}{\sqrt{3}} \hat{I}_+ - \frac{\tilde{R}_- R_2}{\sqrt{3}} \hat{I}_- & -\left( \tilde{R}_+^2 - \frac{R_2^2}{4} \right) \hat{I}_z \end{pmatrix}. \tag{16}$$

Matrix elements $V_{iMM'mm'}$ are then calculated from Eqs.(14, 16) using well-known matrices of raising (lowering) spin operators $\hat{I}_+$ ($\hat{I}_-$) and taking co-ordinate functions of the hole at the nucleus position $\vec{r}_i$. The calculated matrix elements go to either Eq. (9) or Eq. (10), depending on the value of the external magnetic field $B$ as discussed above; then Eq. (12) gives the kinetic equations for $\langle I_z(R) \rangle$. In what follows we consider for simplicity the case of GaAs, where all the nuclei possess spin $3/2$.

The resulting equations connect, for each of the two principle directions, the mean nuclear spin at a distance $R$ from the acceptor ($\langle I_z(R) \rangle$) and the mean non-equilibrium angular momentum of the acceptor-bound hole $\langle \delta \vec{F} \rangle$. For $\vec{R} \parallel \langle \delta \vec{F} \rangle$ ($\theta = 0$) we proceed from Eq (14) and obtain

$$\frac{d\langle I_z \rangle}{dt} \propto \frac{20}{3} R_+^2 \left( R_-^2 - \frac{2}{3} R_+^2 \right) \left( \langle \delta F \rangle - \langle I_z \rangle \right) \tag{17}$$

(both for the weak and strong field regime), i.e., generation and relaxation of the nuclear spin have the same dependence on $R$.

For $\vec{R} \perp \langle \delta \vec{F} \rangle$ ($\theta = \pi/2$) we proceed from the Eq. (16) and obtain

$$\frac{d\langle I_z \rangle}{dt} \propto 10 \langle \delta F \rangle \left[ \frac{2}{3} R_0^4 - \frac{3}{4} R_2^4 + \frac{4}{9} \tilde{R}_-^4 + 3\tilde{R}_+^2 R_2^2 - \frac{5}{3} \tilde{R}_-^2 R_2^2 \right] - 10 \langle I_z \rangle \left[ \frac{2}{3} R_0^4 + \frac{1}{4} R_2^4 + \frac{4}{9} \tilde{R}_-^4 + \tilde{R}_+^2 R_2^2 + \frac{5}{3} \tilde{R}_-^2 R_2^2 \right] \tag{18}$$

for the weak-field regime and

$$\frac{d\langle I_z \rangle}{dt} \propto 10 \langle \delta F \rangle \left[ \frac{2}{3} R_0^4 - \frac{3}{36} R_2^4 + \frac{4}{9} \tilde{R}_-^4 + \frac{3\tilde{R}_+^2 R_2^2}{9} - \frac{5}{3} \tilde{R}_-^2 R_2^2 \right] - 10 \langle I_z \rangle \left[ \frac{2}{3} R_0^4 + \frac{1}{36} R_2^4 + \frac{4}{9} \tilde{R}_-^4 + \frac{\tilde{R}_+^2 R_2^2}{9} + \frac{5}{3} \tilde{R}_-^2 R_2^2 \right] \tag{19}$$

for the strong-field regime. One can see that in the transversal direction, radial dependences of the generation and relaxation of the nuclear spin appear to be different. The specific behavior of the radial dependences for the two principle directions and a parameter set of GaAs are depicted in Fig. 1 (a) for the weak-field regime and in Fig. 2 (a) for the strong-field regime.

### 3.2 Steady-state distribution of the mean nuclear spin

Consider the situation where an external action (e.g., the optical pumping) maintains a stationary value of $\langle \delta F \rangle$. This establishes a stationary spatial distribution of the mean nuclear spin. In order to find it, we use the symmetry considerations of Section 2. In the steady state, the generation and recombination fluxes should balance each other. Equating the right-hand sides of Eqs. (6) and (7), one obtains

$$\langle I_z(R, \theta) \rangle = \frac{\mathcal{A}(R) + \mathcal{B}(R) R^2 \cos^2 \theta}{\tilde{\mathcal{A}}(R) + \tilde{\mathcal{B}}(R) R^2 \cos^2 \theta} \langle \delta F \rangle . \tag{20}$$

Comparing Eq. (20) with either Eqs. (17) and (18) (for the weak-field regime) or Eqs. (17) and (19) (for the strong-field regime), one obtains explicit expressions for $\mathcal{A}, \mathcal{B}, \tilde{\mathcal{A}}, \tilde{\mathcal{B}}$.

For the weak-field regime ($B \gamma_h \tau_c \ll 1$)

$$\mathcal{A} = 10\left[\frac{2}{3}R_0^4 - \frac{3}{4}R_2^4 + \frac{4}{9}\tilde{R}_-^4 + 3\tilde{R}_+^2 R_2^2 - \frac{5}{3}\tilde{R}_-^2 R_2^2\right],$$

$$\mathcal{B}R^2 = \frac{20}{3}R_+^2\left(R_-^2 - \frac{2}{3}R_+^2\right) - 10\left[\frac{2}{3}R_0^4 - \frac{3}{4}R_2^4 + \frac{4}{9}\tilde{R}_-^4 + 3\tilde{R}_+^2 R_2^2 - \frac{5}{3}\tilde{R}_-^2 R_2^2\right],$$

$$\tilde{\mathcal{A}} = 10\left[\frac{2}{3}R_0^4 + \frac{1}{4}R_2^4 + \frac{4}{9}\tilde{R}_-^4 + \tilde{R}_+^2 R_2^2 + \frac{5}{3}\tilde{R}_-^2 R_2^2\right],$$

$$\tilde{\mathcal{B}}R^2 = \frac{20}{3}R_+^2\left(R_-^2 - \frac{2}{3}R_+^2\right) - 10\left[\frac{2}{3}R_0^4 + \frac{1}{4}R_2^4 + \frac{4}{9}\tilde{R}_-^4 + \tilde{R}_+^2 R_2^2 + \frac{5}{3}\tilde{R}_-^2 R_2^2\right]. \tag{21}$$

For the strong-field regime ($B\gamma_h \tau_c \gg 1$)

$$\mathcal{A} = 10\left[\frac{2}{3}R_0^4 - \frac{3}{36}R_2^4 + \frac{4}{9}\tilde{R}_-^4 + \frac{3\tilde{R}_+^2 R_2^2}{9} - \frac{5}{3}\tilde{R}_-^2 R_2^2\right],$$

$$\mathcal{B}R^2 = \frac{20}{3}R_+^2\left(R_-^2 - \frac{2}{3}R_+^2\right) - 10\left[\frac{2}{3}R_0^4 - \frac{3}{36}R_2^4 + \frac{4}{9}\tilde{R}_-^4 + \frac{3\tilde{R}_+^2 R_2^2}{9} - \frac{5}{3}\tilde{R}_-^2 R_2^2\right],$$

$$\tilde{\mathcal{A}} = 10\left[\frac{2}{3}R_0^4 + \frac{1}{36}R_2^4 + \frac{4}{9}\tilde{R}_-^4 + \frac{\tilde{R}_+^2 R_2^2}{9} + \frac{5}{3}\tilde{R}_-^2 R_2^2\right],$$

$$\tilde{\mathcal{B}}R^2 = \frac{20}{3}R_+^2\left(R_-^2 - \frac{2}{3}R_+^2\right) - 10\left[\frac{2}{3}R_0^4 + \frac{1}{36}R_2^4 + \frac{4}{9}\tilde{R}_-^4 + \frac{\tilde{R}_+^2 R_2^2}{9} + \frac{5}{3}\tilde{R}_-^2 R_2^2\right]. \tag{22}$$

Particularly, in the longitudinal direction, for both regimes,

$$\langle I_z \rangle = \langle \delta F \rangle, \tag{23}$$

i.e., the nuclear polarization does not depend on $R$ at all. Certainly, this result is valid only up to distances at which one can neglect other channels of the relaxation of the nuclear spin than the interaction with the hole. In the transverse direction

$$\langle I_z \rangle = \langle \delta F \rangle \frac{\left[\frac{2}{3}R_0^4 - \frac{3}{4}R_2^4 + \frac{4}{9}\tilde{R}_-^4 + 3\tilde{R}_+^2 R_2^2 - \frac{5}{3}\tilde{R}_-^2 R_2^2\right]}{\left[\frac{2}{3}R_0^4 + \frac{1}{4}R_2^4 + \frac{4}{9}\tilde{R}_-^4 + \tilde{R}_+^2 R_2^2 + \frac{5}{3}\tilde{R}_-^2 R_2^2\right]} \tag{24}$$

in the weak-field regime and

$$\langle I_z \rangle = \langle \delta F \rangle \frac{\left[\frac{2}{3}R_0^4 - \frac{3}{36}R_2^4 + \frac{4}{9}\tilde{R}_-^4 + \frac{3\tilde{R}_+^2 R_2^2}{9} - \frac{5}{3}\tilde{R}_-^2 R_2^2\right]}{\left[\frac{2}{3}R_0^4 + \frac{1}{36}R_2^4 + \frac{4}{9}\tilde{R}_-^4 + \frac{\tilde{R}_+^2 R_2^2}{9} + \frac{5}{3}\tilde{R}_-^2 R_2^2\right]} \tag{25}$$

in the strong-field regime.

Maps of the steady-state nuclear polarization are presented in Fig.1 (b) and Fig. 2 (b). Very close to the acceptor, the nuclear polarization in distributed almost spherically-symmetric, but at a distance the distribution becomes anisotropic.

## 3.3 Splitting of the acceptor sublevels via the interaction with polarized nuclei

Suppose the polarized acceptor-bound holes have created the steady-state nuclear polarization, as given by Eq. (20). Let us calculate the energy splitting of the acceptor ground state by means of its interaction with the polarized nuclei. We make use of the complete expressions of the acceptor wave functions Eq. (B1) and the matrix Eq. (A9), but keep only the contributions containing $\hat{I}_z$, since the contributions to the energy from the terms containing $\hat{I}_+$ and $\hat{I}_-$ are equal to zero on average.
This way, one obtains

$$\langle \Psi_M | \hat{H}_{dd} | \Psi_{M'} \rangle = \frac{32}{15}\pi \frac{\mu_B \mu_I}{I} \wp \hat{I}_z \begin{pmatrix} \left(A_+^2 + \frac{C^2}{3} - \frac{D^2}{3}\right) & -\left(A_+ C + \frac{A_- C}{3}\right)e^{-i\phi} & -\left(A_+ D - \frac{A_- D}{3}\right)e^{-2i\phi} & \frac{2}{3}CDe^{-3i\phi} \\ -\left(A_+ C + \frac{A_- C}{3}\right)e^{+i\phi} & \left(\frac{A_-^2}{3} + C^2 - D^2\right) & 2CDe^{-i\phi} & \left(A_+ D - \frac{A_- D}{3}\right)e^{-2i\phi} \\ -\left(A_+ D - \frac{A_- D}{3}\right)e^{+2i\phi} & 2CDe^{+i\phi} & -\left(\frac{A_-^2}{3} + C^2 - D^2\right) & -\left(A_+ C + \frac{A_- C}{3}\right)e^{-i\phi} \\ \frac{2}{3}CDe^{+3i\phi} & \left(A_+ D - \frac{A_- D}{3}\right)e^{+2i\phi} & -\left(A_+ C + \frac{A_- C}{3}\right)e^{+i\phi} & -\left(A_+^2 + \frac{C^2}{3} - \frac{D^2}{3}\right) \end{pmatrix}$$

(26)

where

$$A_+ = \frac{R_0}{\sqrt{4\pi}} + \frac{R_2}{\sqrt{16\pi}}(1 - 3\cos^2\theta), \quad A_- = \frac{R_0}{\sqrt{4\pi}} - \frac{R_2}{\sqrt{16\pi}}(1 - 3\cos^2\theta), \tag{27}$$

$$C = \sqrt{\frac{3}{4\pi}} R_2 \sin\theta \cos\theta, \quad D = \sqrt{\frac{3}{16\pi}} R_2 \sin^2\theta.$$

Then one should add together the contributions from all the nuclei, i.e., in fact, integrate over $R, \theta, \phi$ with the spatially-inhomogeneous nuclear polarization Eq. (20). As a result, all the off-diagonal matrix elements Eq. (26) turn to zero after integration over $\phi$. This means that the projection of the total angular momentum of the acceptor-bound hole remains a good quantum number for classification of the split sublevels. However, the pattern of the split levels appears to be somewhat unusual – not like the splitting caused by an external magnetic field or by homogeneously-polarized nuclei. Averaging of the diagonal matrix elements Eq. (26) reveals that the separation between the sublevels $M = \pm 3/2$ and that between the sublevels $M = \pm 1/2$ are not in the customary ratio 3:1. In other words, the splitting of the acceptor ground state via the interaction with the nuclear polarization appears to be non-equidistant, and, strictly speaking, the action of the nuclei onto the acceptor center can not be described in terms of the Overhauser field. For the specific parameters of GaAs, the ratio of the spin splittings of holes $M = \pm 3/2$ and of holes $M = \pm 1/2$ equals 3.43 in the weak-field regime ($B\gamma_h\tau_c \ll 1$) and 2.73 in the strong-field regime ($B\gamma_h\tau_c \gg 1$).

**4. Possible ways of experimental observation of the dynamical polarization of nuclei by the acceptor-bound holes**

The most efficient method of the dynamical polarization of nuclei via the interaction with the conduction-band electrons – creation of a non-equilibrium orientation of the electron spins by means of the optical pumping[16] or spin injection.[19,20] For holes, this method is possible in principle but hindered by the fast spin relaxation of free holes in the valence band.[16] One possible way to surmount this difficulty is using resonant optical pumping of the acceptor levels by circularly polarized light. Another possible approach involves the Overhauser effect, i.e., creation of the non-equilibrium angular momentum of holes by means of their *depolarization*, performed either by a non-polarized light or by microwaves.[21] In such an experiment, one ought to apply a rather strong magnetic field and maintain the sample at a low temperature, in order to achieve a significant *equilibrium* value of the mean angular momentum of the acceptor-bound holes.

In any of the proposed experiments, the non-equidistant splitting of the acceptor levels will become a key distinctive feature. The spin-flip Raman scattering and the electron spin resonance are adequate experimental techniques for observation of the effect. The magnetic resonance technique combines well with the dynamical nuclear polarization through the Overhauser effect, since the latter implies a noticeable equilibrium polarization of holes whose resonant breakdown by the microwave field can be conveniently detected through the change of the circular polarization degree of the luminescence (in fact, a kind of optically detected magnetic resonance).

**5. Conclusions**

We have addressed the problem of acceptor-bound holes in a cubic semiconductor which polarize spins of the crystal lattice nuclei by means of the dipole-dipole interaction. We have shown that this interaction is scalar, though, in contrast with the hyperfine interaction of conduction-band electrons, it does not have a contact form. We have considered the experimental design with an external magnetic field applied, which is stronger than the local dipole-dipole fields of the neighboring nuclei, and a non-equilibrium polarization of the angular momentum of acceptor-bound holes created along this field by some external force like optical pumping or microwave power. We calculated coordinate dependences of the rate of polarization of the nuclear spin system and of the rate of relaxation of non-equilibrium nuclear spins by means of their interaction with the hole. Two regimes were shown to result from the model, depending on the relation between the Zeeman splitting of hole sublevels, driven by the value of the applied magnetic field, and the homogeneous width of these sublevels, controlled by the spin correlation time of the hole. For both the weak-field and the strong-field regimes, we have found steady-state distributions (maps) of the nuclear spin polarization as well as splitting patterns of the sublevels of the acceptor state.

The steady-state spatial distribution of the nuclear polarization turns out to be essentially anisotropic, including distances comparable to the localization radius of the acceptor-bound hole. This anisotropy results in the non-equidistance of sublevels of the net angular momentum of the acceptor interacting with spin-polarized nuclei. The non-equidistance can experimentally manifest itself, for instance, in spin-flip Raman scattering, thus being a key observable in the problem. Hence, it is important to make sure that other channels of relaxation of the nuclear spin system (spin diffusion in the first place) are not able to smear out the anisotropy in the nuclear spin distribution and to destroy the non-equidistance effect.

Let us evaluate the characteristic time of the diffusive relaxation of inhomogeneities of the nuclear polarization on the spatial scale corresponding to $R = 5$ (nuclear polarization within this radius makes a dominant contribution to the acceptor sublevel splitting). It comprises about 60 Å for GaAs. Adopting, for the diffusion coefficient of the nuclear spin, the estimate $D \sim 10^{-13}$ cm$^2$/s,[16] we obtain the diffusion time $\tau_D \approx 3.6$ s. Looking for the characteristic time of the polarization of nuclei by the holes, we make use of the evaluation conducted in Ref.[16] for donor-bound electrons and assume the hyperfine constant ten times smaller that that for electrons[2–4] while the correlation time of the same scale. Taking into account a stronger localization of the hole on the acceptor (characteristic radius 20 Å) and a realistic profile of the hole radial functions (see insets in Figs. 1,2), we obtain for the same distance (60 Å away from the acceptor) a typical time of the spin transfer by the hole to nuclear system: $\tau_h \sim 0.3$ s. Inequality $\tau_D \gg \tau_h$ justifies the model considered in the present paper.

**Acknowledgements**

We gratefully acknowledge helpful advices from A.V. Rodina regarding the parameter choice for the variational wave functions of the acceptor state.
This paper was partially supported by EU FET project SPANGL4Q and by the Russian Ministry of Education and Science (contract No. 11.G34.31.0067 with SPbSU and leading scientist A.V. Kavokin). AK acknowledges support from the Dmitry Zimin "Dynasty" Foundation.

**Appendix A. The Hamiltonian of the magneto-dipole interaction in the matrix representation**

The Hamiltonian of the magnetic dipole-dipole interaction has the form[21]

$$\hat{H}_{dd} = \frac{2\mu_B \mu_I}{I\rho^3} \hat{\vec{I}} \cdot \left( \hat{\vec{l}} - \hat{\vec{s}} + 3\frac{\vec{\rho}(\hat{\vec{s}} \cdot \vec{\rho})}{\rho^2} \right) \qquad (A1)$$

where $\hat{\vec{l}}$, $\hat{\vec{s}}$ and $\vec{\rho}$ are the orbital moment operator, the spin operator and position vector of the electron, $\hat{\vec{I}}$ is the operator of the nuclear spin, $\mu_B$ and $\mu_I$ are the Bohr magneton and the nuclear magneton, respectively.

Let us calculate the matrix elements of this Hamiltonian

$$\langle u_m | \hat{H}_{dd} | u_{m'} \rangle = \int d^3 r \cdot u_m^* \hat{H}_{dd} u_{m'}, \quad m = \pm 3/2, \pm 1/2 \qquad (A2)$$

with the Bloch functions of the complex valence band[22]

$$u_{+3/2} = -\frac{u(\rho)}{\sqrt{2}}(x+iy)\uparrow, \qquad (A3)$$

$$u_{+1/2} = \frac{u(\rho)}{\sqrt{3}}\left(-\frac{1}{\sqrt{2}}(x+iy)\downarrow + \sqrt{2}z\uparrow\right),$$

$$u_{-1/2} = \frac{u(\rho)}{\sqrt{3}}\left(\frac{1}{\sqrt{2}}(x-iy)\uparrow + \sqrt{2}z\downarrow\right),$$

$$u_{-3/2} = \frac{u(\rho)}{\sqrt{2}}(x-iy)\downarrow;$$

here $x, y, z$ are orbital functions which are transformed as the respective coordinates, up and down arrows indicate contributions from the corresponding spin components.

The components of the orbital moment operator $\hat{l}$ (with $l=1$) are[23] matrices $3\times 3$

$$\hat{l}_x = \begin{pmatrix} 0 & 1/\sqrt{2} & 0 \\ 1/\sqrt{2} & 0 & 1/\sqrt{2} \\ 0 & 1/\sqrt{2} & 0 \end{pmatrix}, \hat{l}_y = \begin{pmatrix} 0 & -i/\sqrt{2} & 0 \\ i/\sqrt{2} & 0 & -i/\sqrt{2} \\ 0 & i/\sqrt{2} & 0 \end{pmatrix}, \hat{l}_z = \begin{pmatrix} 1 & 0 & 0 \\ 0 & 0 & 0 \\ 0 & 0 & -1 \end{pmatrix}, \qquad (A4)$$

while the components of the spin operator $\hat{\vec{s}}$ are Pauli matrices multiplied by $1/2$:

$$\hat{s}_x = \begin{pmatrix} 0 & 1/2 \\ 1/2 & 0 \end{pmatrix}, \hat{s}_y = \begin{pmatrix} 0 & -i/2 \\ i/2 & 0 \end{pmatrix}, \hat{s}_z = \begin{pmatrix} 1/2 & 0 \\ 0 & -1/2 \end{pmatrix}. \qquad (A5)$$

The matrices of the orbital moment are written in the basis of the spherical functions

$$Y_{1,0} = i\sqrt{\frac{3}{4\pi}}\cos\vartheta, \quad Y_{1,\pm1} = \mp i\sqrt{\frac{3}{8\pi}}\sin\vartheta \cdot e^{\pm i\varphi}, \tag{A6}$$

so the Bloch functions Eq. (3) should be transformed to the same basis. One obtains

$$u_{+3/2} = -\frac{2i\sqrt{\pi}}{\sqrt{3}}\rho u(\rho)Y_{1,+1}\uparrow, \tag{A7}$$

$$u_{+1/2} = -\frac{2i\sqrt{\pi}}{3}\rho u(\rho)Y_{1,+1}\downarrow - \frac{i\sqrt{8\pi}}{3}\rho u(\rho)Y_{1,0}\uparrow,$$

$$u_{-1/2} = -\frac{2i\sqrt{\pi}}{3}\rho u(\rho)Y_{1,-1}\uparrow - \frac{i\sqrt{8\pi}}{3}\rho u(\rho)Y_{1,0}\downarrow,$$

$$u_{-3/2} = -\frac{2i\sqrt{\pi}}{\sqrt{3}}\rho u(\rho)Y_{1,-1}\downarrow.$$

The calculation to follow will require the knowledge of the result of the action of the angular momentum matrices on the three spherical harmonics: $Y_{1,0}$, $Y_{1,\pm1}$. It is presented in Table 1. Now one obtains the matrix elements Eq. (A2), operating with the Bloch functions Eqs. (A7) and using the relations

$$\frac{x^2}{\rho^2} = \sin^2\vartheta\cos^2\varphi, \quad \frac{y^2}{\rho^2} = \sin^2\vartheta\sin^2\varphi, \quad \frac{z^2}{\rho^2} = \cos^2\vartheta, \tag{A8}$$

$$\frac{xy}{\rho^2} = \sin^2\vartheta\sin\varphi\cos\varphi, \quad \frac{xz}{\rho^2} = \sin\vartheta\cos\vartheta\cos\varphi, \quad \frac{yz}{\rho^2} = \sin\vartheta\cos\vartheta\sin\varphi.$$

Integration over $\rho$, $\vartheta$ and $\varphi$ leads to the following final result:

$$\langle u_m|\hat{H}_{dd}|u_{m'}\rangle = \frac{64}{45}\pi\frac{\mu_B\mu_I}{I}\wp\begin{pmatrix} \frac{3}{2}\hat{I}_z & \frac{\sqrt{3}}{2}\hat{I}_- & 0 & 0 \\ \frac{\sqrt{3}}{2}\hat{I}_+ & \frac{1}{2}\hat{I}_z & \hat{I}_- & 0 \\ 0 & \hat{I}_+ & -\frac{1}{2}\hat{I}_z & \frac{\sqrt{3}}{2}\hat{I}_- \\ 0 & 0 & \frac{\sqrt{3}}{2}\hat{I}_+ & -\frac{3}{2}\hat{I}_z \end{pmatrix}, \tag{A9}$$

where operators $\hat{I}_z$, $\hat{I}_+ = \hat{I}_x + i\hat{I}_y$, $\hat{I}_- = \hat{I}_x - i\hat{I}_y$ act on the nuclear spins, while the radial integral $\wp = \int_0^\infty u^2(\rho)\rho d\rho$, determined by the specific radial function $u(\rho)$, can be found either from comparison with experimental data or from first-principle numerical calculations.

One can see that the matrix in Eq. (A9) represents the operator $\vec{J}\cdot\vec{I}$, which confirms symmetry considerations that have lead us to Eq. (2) of the main text.

## Appendix B. Wave functions of the acceptor-bound hole in the complex valence band

Wave functions of the hole with the total angular moment $3/2$, localized at a shallow acceptor in a semiconductor with a diamond-like crystal lattice and a large spin-orbit splitting of the valence band, read[10]

$$\psi_{+3/2} = R_0 Y_{0,0} u_{+3/2} + \frac{1}{\sqrt{5}} R_2 Y_{2,0} u_{+3/2} - \frac{2}{\sqrt{10}} R_2 Y_{2,+1} u_{+1/2} + \frac{2}{\sqrt{10}} R_2 Y_{2,+2} u_{-1/2}, \qquad (B1)$$

$$\psi_{+1/2} = R_0 Y_{0,0} u_{+1/2} + \frac{2}{\sqrt{10}} R_2 Y_{2,-1} u_{+3/2} - \frac{1}{\sqrt{5}} R_2 Y_{2,0} u_{+1/2} + \frac{2}{\sqrt{10}} R_2 Y_{2,+2} u_{-3/2},$$

$$\psi_{-1/2} = R_0 Y_{0,0} u_{-1/2} + \frac{2}{\sqrt{10}} R_2 Y_{2,-2} u_{+3/2} - \frac{1}{\sqrt{5}} R_2 Y_{2,0} u_{-1/2} + \frac{2}{\sqrt{10}} R_2 Y_{2,+1} u_{-3/2},$$

$$\psi_{-3/2} = R_0 Y_{0,0} u_{-3/2} + \frac{1}{\sqrt{5}} R_2 Y_{2,0} u_{-3/2} + \frac{2}{\sqrt{10}} R_2 Y_{2,-2} u_{+1/2} - \frac{2}{\sqrt{10}} R_2 Y_{2,-1} u_{-1/2}.$$

They depend on two radial functions $R_0 = R_0(R)$ and $R_2 = R_2(R)$, where $R$ stands for the dimensionless distance from the acceptor center (the same as $r$ but measured in the units of Bohr radius with the heavy-hole mass, ~12.5 Å for GaAs). At an arbitrary ratio of the light-hole mass and the heavy-hole mass, the functions $R_0$ and $R_2$ have no exact analytical expression. As shown by Rodina and Averkiev,[24,25] the following variational approximation of these functions works very well:

$$R_0(R) = \sqrt{2}\left[1+(\zeta\beta)^{3/2}\right]^{-1/2}\left[\tilde{R}_0(\alpha R)+(\zeta\beta)^{3/2}\tilde{R}_0(\alpha\sqrt{\zeta\beta}R)\right], \qquad (B2)$$

$$R_2(R) = \sqrt{2}\left[1+(\zeta\beta)^{3/2}\right]^{-1/2}\left[\tilde{R}_2(\alpha R)-(\zeta\beta)^{3/2}\tilde{R}_2(\alpha\sqrt{\zeta\beta}R)\right],$$

where

$$\tilde{R}_0(x) = \alpha^{3/2} e^{-x}, \qquad (B3)$$

$$\tilde{R}_2(x) = \alpha^{3/2}\left[\frac{6}{x^3} - e^{-x}\left(1+\frac{3}{x}+\frac{6}{x^2}+\frac{6}{x^3}\right)\right],$$

while the values of the variational parameters $\alpha$ and $\zeta$ as functions of the light-to-heavy mass ratio $\beta$ can be found by minimization of the total energy.[24,25] For instance, for the case of GaAs, proceeding from the Luttinger parameters set of Ref. 26 ($\gamma_1 = 6.98$, $\gamma_2 = 2.06$, $\gamma_3 = 2.93$), one obtains $\gamma = (2\gamma_2 + 3\gamma_3)/5 = 2.58$, $\beta = (\gamma_1 - 2\gamma)/(\gamma_1 + 2\gamma) = 0.15$ and finds the corresponding values $\alpha = 0.707$ and $\zeta = 1.69$.

**References**

[1] G. Lampel. Phys. Rev. Lett. **20**, 491 (1968)

[2] E. I. Gryncharova and V. I. Perel. Fiz. Tekhn. Poluprovodn., **11**, 1697 (1977) [Sov. Phys. Semicond. **11**, 997 (1977)]

[3] J. Fischer, W. A. Coish, D. V. Bulaev, and D. Loss. Phys. Rev. B **78**, 155329 (2008)

[4] C. Testelin, F. Bernardot, B. Eble, and M. Chamarro. Phys. Rev. B **79**, 195440 (2009)

[5] P. Fallahi, S. T. Yılmaz, and A. Imamoğlu. Phys. Rev. Lett. **105**, 257402 (2010)

[6] E. A. Chekhovich, A. B. Krysa, M. S. Skolnick, and A. I. Tartakovskii. Phys. Rev. Lett. **106**, 027402 (2011)

[7] E. A. Chekhovich, M. M. Glazov, A. B. Krysa, M. Hopkinson, P. Senellart, A. Lemaitre, M. S. Skolnick, and A. I. Tartakovskii. Nature Phys. 9, 74 (2013)

[8] Wen Yang and L. J. Sham. Phys. Rev. B **85**, 235319 (2012)

[9] N. O. Lipari and A. Baldereschi. Phys. Rev. Lett. **25**, 1660 (1970)

[10] B. L. Gel'mont and M. I. Dyakonov. Fiz. Tekh. Poluprovodn. **5**, 2191 (1971) [Sov. Phys. Semicond. **5**, 1905 (1972)]

[11] A. V. Malyshev, I. A. Merkulov, and A. V. Rodina. Phys. Rev. B **55**, 4388 (1997)

[12] Al. L. Efros, M. Rosen, M. Kuno, M. Nirmal, D. J. Norris, and M. Bawendi. Phys. Rev. B **54**, 4843 (1996 )

[13] A. V. Rodina and Al. L. Efros. Phys. Rev. B **82**, 125324 (2010)

[14] Y. F. Berkovskaya, E.M. Vakhabova, B.L. Gelmont, I.A. Merkulov. Zh. Eksp. Teor. Phys. **94**, 183 (1988) [Sov. Phys. JETP **67**, 750 (1988)]

[15] I. A. Merkulov and A. V. Rodina. "Exchange interaction between carriers and magnetic ions in quantum size heterostructures", in: "Introduction to the Physics of Diluted Magnetic Semiconductors", eds. J. Kossut and J. Gaj, Springer Series in Materials Science, Vol. 144, pp. 65-101 (2010)

[16] *Optical Orientation*, edited by F. Meyer and B. P. Zakharchenya (North-Holland, Amsterdam, 1984)

[17] V. K. Kalevich, K. V. Kavokin, I. A. Merkulov. "Dynamic nuclear polarization and nuclear fields", in "Spin physics in semiconductors", ed. by M.I.Dyakonov. Springer, 2008.

[18] M. I. Dyakonov and V. I. Perel'. Zh. Eksp. Teor. Phys. **63**, 1883 (1972) [Sov. Phys. JETP **36**, 995 (1973)]

[19] M. Johnson. Appl. Phys. Lett. **77**, 1680 (2000)

[20] R. J. Epstein, J. Stephens, M. Hanson, Y. Chye, A. C. Gossard, P. M. Petroff, and D. D. Awschalom. Phys. Rev. B **68**, 041305 (2003)

[21] A. Abragam. *The principles of Nuclear Magnetism* (Oxford University Press, Ney York, 1961)

[22] E. L. Ivchenko, *Optical Spectroscopy of Semiconductor Nanostructures* (Alpha Science, Harrow, UK, 2005)

[23] E. L. Ivchenko and G. E. Pikus, *Superlattices and Other Heterostructures: Symmetry and Optical Phenomena*, Springer Series in Solid State Sciences Vol. **110** (Springer-Verlag, Berlin, 1995)

[24] A. V. Rodina. Sol. State Comm. **85**, 23 (1993)

[25] N. S. Averkiev and A. V. Rodina. Fiz. Tverdogo Tela **35**, 1051 (1993) [Phys. Solid State **35**, 538 (1993)]

[26] I. Vurgaftman, J.R. Meyer, L.R. Ram-Mohan. J. Appl. Phys. 89, 5815 (2001)

Table 1. Action of the orbital moment components onto the spherical harmonics

| $\hat{l}_i$ | $\hat{l}_i\|Y_{1,+1}\rangle$ | $\hat{l}_i\|Y_{1,0}\rangle$ | $\hat{l}_i\|Y_{1,-1}\rangle$ |
|---|---|---|---|
| $\hat{l}_x$ | $\dfrac{1}{\sqrt{2}}Y_{1,0}$ | $\dfrac{1}{\sqrt{2}}(Y_{1,+1}+Y_{1,-1})$ | $\dfrac{1}{\sqrt{2}}Y_{1,0}$ |
| $\hat{l}_y$ | $\dfrac{i}{\sqrt{2}}Y_{1,0}$ | $-\dfrac{i}{\sqrt{2}}(Y_{1,+1}-Y_{1,-1})$ | $-\dfrac{i}{\sqrt{2}}Y_{1,0}$ |
| $\hat{l}_z$ | $Y_{1,+1}$ | $0$ | $-Y_{1,-1}$ |

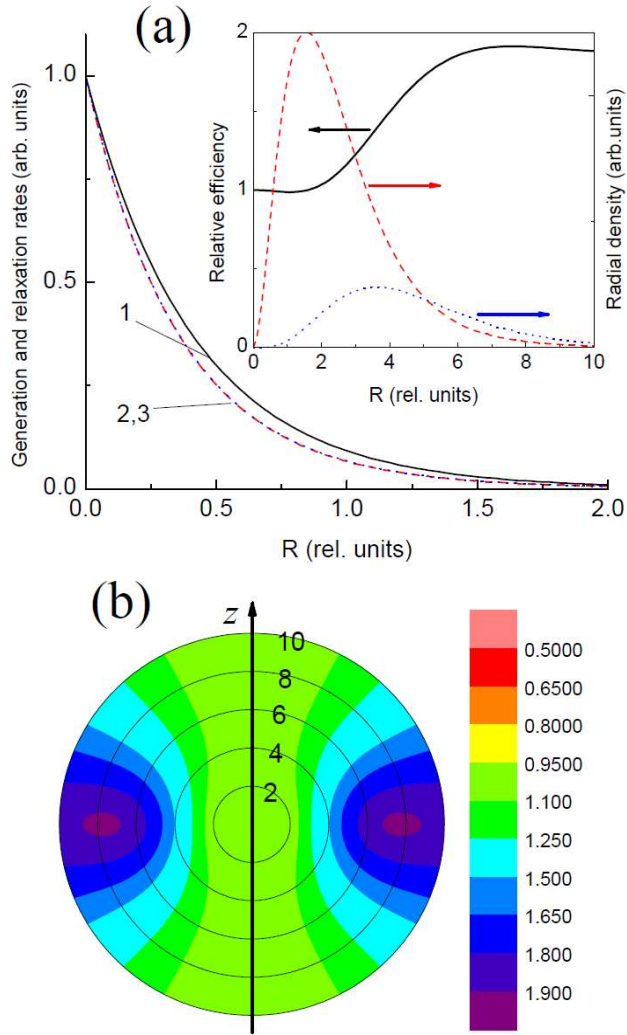

Figure 1. Hyperfine interaction of the acceptor-bound hole and nuclear polarization in the weak-field regime (see text). (a) Radial dependences of the rates of generation and relaxation of the nuclear spin in the vicinity of the acceptor center: along the $Z$ direction (curve 1; dependences for generation and relaxation coincide) and in the perpendicular direction (curves 2 and 3 almost merge in the scale of the figure). $R=1$ corresponds to the length 12.5 Å for GaAs (effective Bohr radius calculated with the heavy-hole mass). The inset shows how the ratio of the generation and relaxation efficiencies behaves at larger distances from the acceptor (solid curve). For comparison, the radial densities of the wave functions $R_0^2 R^2$ (dashed line) and $R_2^2 R^2$ (dotted line) are shown. One can see that the efficiency ratio deviates from unity together with the onset of the $R_2$ orbital. (b) Map of the steady-state nuclear polarization in the vicinity of the acceptor (meridianal cross-section). Numbers at the concentric circles indicate distances $R$ from the acceptor center.

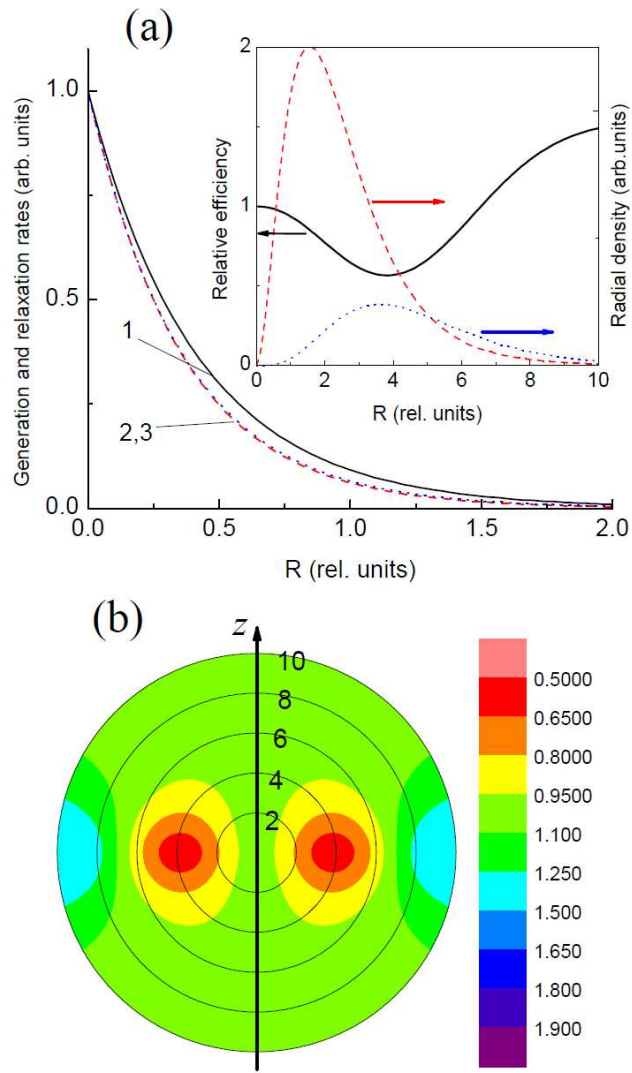

Figure 2. The same as in Fig.1 but for the strong-field regime (see text).